\font\msamstex=msam10
\def\gtrsim {\,\mbox{{\msamstex \char 38}}\,}
\def\lesssim{\,\mbox{{\msamstex \char 46}}\,}
\documentstyle[
aps,prb]{revtex}

\begin{document}
\draft
\title{Improved Gaussian self--consistent method  ---
Applications to homopolymers with different architectures in dilute solution}
\author{Edward G.~Timoshenko\thanks{Corresponding author. 
Web: http://darkstar.ucd.ie; 
E-mail: Edward.Timoshenko@ucd.ie}
}
\address{
Theory and Computation Group,
Department of Chemistry, University College Dublin,
Belfield, Dublin 4, Ireland}

\author{Yuri A.~Kuznetsov}
\address{
Centre for High Performance Computing Applications, 
University College Dublin,
Belfield, Dublin 4, Ireland}

\date{\today}
\maketitle

\begin{abstract}

A version of the Gaussian self--consistent (GSC) method, 
which avoids the use of the Edwards' virial expansion, is
presented. Instead, the mean energy is evaluated directly
via a convolution of the attractive part of the pair--wise non--bonded
potential with the Gaussian trial radial
distribution function. The hard sphere repulsion is
taken into account via a suitably generalised Carnahan--Starling term.
Comparison of the mean--squared inter--monomer distances and radius
of gyration, as well as of the mean energy, between
the results from the GSC calculations and Monte Carlo (MC) simulation in
continuous space are made across the coil--to--globule
transition for isolated ring, open and star
homopolymers of varied lengths and flexibility. Importantly, both
techniques utilise the same polymer model so that the data points
could be directly superimposed. A surprisingly good overall agreement
is found between these GSC and MC results. Caveats of the Gaussian
technique and ways for going beyond it are also discussed.
\end{abstract}

\vskip 1cm

\section{Introduction}
\label{sec:intro}

The Gaussian self--consistent (GSC) method represents a
quite general, albeit relatively simple, theoretical framework
for studying the equilibrium, dynamics and kinetics
of conformational changes in polymer solutions. One of its
most attractive features is that it can be applied to
virtually any type of heteropolymer
\cite{ConfTra,Ganazzoli-98,Ganazzoli-00}, 
with any specific interaction terms involving e.g.\ chain stiffness 
\cite{Torus,Ganazzoli} or
charges \cite{Netz}. The connectivity of the chain
can be arbitrary: an open polymer \cite{KineticLaws}, 
a ring \cite{GSC1st}, a star \cite{CopStar}, or a dendrimer
\cite{Ganazzoli-dendr}, 
whether in the limit of a single chain \cite{QzMess}
or at finite concentrations \cite{Analysis}.
Moreover, not only the radius of gyration of the polymer or 
the total density can be computed, but much finer details of the 
conformational structure, such as
the mean--squared distances between monomers or radial distributions
at a given time, are available.
Such a versatility naturally comes at the cost of certain limitations
and inaccuracies intrinsic to the technique. However, these are well
known and manageable, although often largely overstated in
the popular view. Some of these, such as the Gaussian shape of
the radial distribution function (RDF), are quite unavoidable
at the level of the GSC theory, but others, such as the traditional
use of the Edwards' virial expansion \cite{Edwards,CloizeauxBook}, 
can indeed be surpassed as we shall demonstrate
in the current work. This would permit us for the first time to compare
observables between the predictions of the GSC technique and Monte
Carlo (MC) simulation in continuous space 
based on the same model for given values 
of thermodynamic parameters throughout the range of
the coil--to--globule transition. 

Historically, perhaps the first of the equilibrium versions of
the Gaussian method in application to a single homopolymer chain
were independently proposed by S.~Edwards {\it et al} 
\cite{Edwards-var} and J.~des Cloizeaux \cite{Cloizeaux-var}.
The approach of the work by Edwards \cite{Edwards-var} relied
on the following idea.
One would like to approximate the coordinates of the real homopolymer chain,
$\bbox{X}_i$, governed by the Gibbs distribution
with the exact Hamiltonian $H$, via a `trial' Gaussian chain
with the monomer positions $\bbox{X}_i^{(0)}$, governed by
a simpler trial Hamiltonian $H^{(0)}$.  By assuming that
the quantities $\Delta \bbox{X}_i = \bbox{X}_i - \bbox{X}_i^{(0)}$
and $\Delta H \equiv H -H^{(0)}$ are small and hence by applying the
Taylor expansion to the equilibrium Gibbs averages,
in the first order, one requires that the mean--squared
radius of gyration obeys,
\begin{equation}
\label{Edwards}
\langle R_g^2 \rangle_0 = \langle R_{g}^{(0)\, 2} \rangle_0,\quad
\mbox{or equivalently}\ \ 
\langle R_{g}^{(0)\, 2} H^{(0)} \rangle_0 -
\langle R_{g}^{(0)\, 2} \rangle_0\, \langle H^{(0)} \rangle_0 = 0.
\end{equation}
Such a theory predicts the correct mean--field swelling exponent
$\nu_F = 3/5$ for the repulsive coil, only missing the subtle
renormalisation group correction.
It also gives appropriately $\nu_{\theta} = 1/2$ 
for the ideal coil, and $\nu_g = 1/3$ for the globule.
Moreover, this approach can be further extended to kinetics
by considering time dependent analogues of
Eq. (\ref{Edwards}) \cite{OrlandPitard,Pitard,ConfTra,GSC1st}.

A seemingly more refined theory proposed by des Cloizeaux \cite{Cloizeaux-var}
uses $N-1$ (where $N$ is the degree of polymerisation)
variational variables $\bbox{X}_{q}^{(0)}$, which are
the normal modes of the monomer positions for a ring chain 
$\bbox{X}_{i}^{(0)}$.
This technique is based on the standard Gibbs--Bogoliubov variational
principle with a quadratic diagonal trial
Hamiltonian $H^{(0)}$. The equilibrium corresponds to the minimum 
of the trial free energy, ${\cal A}_{trial} = {\cal A}^{(0)} +
\langle H - H^{(0)} \rangle_0$, where ${\cal A}^{(0)}$
is the free energy associated with $H^{(0)}$. 
Expressed in the Edwards' form the extremum conditions for the trial
free energy are precisely the following $N-1$ equations,
\begin{equation}
\langle \bbox{X}_{q}^2 \rangle_0 = \langle \bbox{X}_{q}^{(0)\, 2} \rangle_0,
\quad \mbox{or equivalently}\ \ 
\langle \bbox{X}_{q}^{(0)\, 2} H^{(0)} \rangle_0 - 
\langle \bbox{X}_{q}^{(0)\, 2} \rangle_0\, \langle H^{(0)} \rangle_0 = 0.
\end{equation}
Unfortunately, this theory, although more accurate for polymers
around the theta--point and indeed for the globule, is known to have
a deficiency in that it predicts the exponent $\nu = 2/3$ instead of
the mean--field value of $\nu_F = 3/5$ for the swollen coil
in the thermodynamic limit $N\to \infty$.
For this reason, some suspicion arose concerning the validity of the
theory by des~Cloizeaux,  hampering
further efforts in this direction.
It should be emphasised, however, that for finite values of $N$
the resulting theory does give quite reasonable numerical
results fairly close to $\nu_F = 3/5$. 
In fact, the numerically fitted value of the swelling exponent tends to be 
somewhere in between the two above theoretical predictions.
In a way, this is not an unusual situation, even for the more elaborate
integral equations theories of molecular fluids used in attempts to deduce
the Flory exponent \cite{Eu,Taylor}.
More importantly, this well known deficiency of the 
theory by des~Cloizeaux becomes irrelevant when considering the
theta--point or the poor solvent conditions.

Therefore, activity on further extending and improving the GSC theory
has persisted, notably in Milan by G.~Allegra and F.~Ganazzoli {\it et al}
\cite{Allegra,Ganazzoli,Ganazzoli-00,Ganazzoli-98,Allegra-93,Raos-94,%
Ganazzoli-dendr}, in Saclay by H.~Orland and T.~Garel {\it et al}
\cite{OrlandPitard,Pitard,Netz}, as well as in our Group in Dublin
\cite{GSC1st,KineticLaws,ConfTra,Torus,QzMess,Analysis},
and some collaborations between these have emerged \cite{CopStar}.
Furthermore, extensions of the GSC theory to kinetics, as well as
to essentially any polymeric system, have been also achieved.
For the former end, one proceeds from the Langevin equation
and approximates the exact stochastic ensemble via a linear trial one.
We refer the reader to our papers in Refs. \onlinecite{ConfTra,QzMess,CopStar}
for the ultimate formulation of the GSC theory in terms of the real
space variables. Such a version of the GSC method avoids 
the limitations of the earlier normal modes
formulations since it can distinguish the frustrated states of
heteropolymers with spontaneously broken kinematic symmetries associated
with the connectivity structure.

Despite the relative success of the most recent form of the GSC method,
the theoretical situation is not as yet fully satisfactory and
further efforts along these lines are required.
Firstly, one would like to be able to use
the actual molecular interaction parameters rather than
the coefficients in the Edwards' virial expansion,
which are somewhat obscure in their meaning.
Secondly, the convergence  of such a virial expansion in the dense
liquid globular state is rather problematic, being at best
that of an asymptotic series.
Thirdly, despite the conventional view, the three--body term
is unable by itself to fully withstand a catastrophic pair--wise
collapse of monomers onto each other in the attractive
regime for heteropolymers.
In Refs. \onlinecite{ConfTra,QzMess} an {\it ad hoc} self--interaction
energy term has been added to tackle this problem, but it is merely
a tweak of the model.
Finally, the effective three--body and higher order virial terms
make the current theory numerically inefficient due to
a scaling law involving a high power in $N$ for the time
expense per step, despite the fact that the original model
only contains two--body interactions.

Therefore, in the current work we shall rid ourselves 
of relying on the virial representation of the Hamiltonian entirely.
Indeed, the exact molecular two--body potential can be averaged
over the Gaussian RDF directly.
However, the Gaussian function does not vanish at the origin,
yielding a divergence from the repulsive part of the Lennard--Jones 
pair--wise potential. One simple alternative is to use the latter
potential with a hard core part, as in Ref. \onlinecite{CorFunc},
in which case one has to cut off the integration at the hard sphere
diameter, or more generally, within the excluded volume area. 

To include the effect of the hard sphere repulsion one can
in principle use any of the standard reference techniques
developed for the hard sphere liquids, e.g. the Percus--Yevick expression
for RDF, or the Carnahan--Starling (CS) 
free energy formula \cite{HansenMacDonald,CarnahanStarling}.
We prefer to use the latter due to its practical simplicity and accuracy
when compared to molecular simulations. The Carnahan--Starling equation,
which was originally derived for a single component hard sphere liquid, has
been naturally extended to mixtures. It is quite popular in
the liquid matter literature at present (see e.g. Ref. \onlinecite{Talanquer}). 
Its extension to polymeric fluids is somewhat less obvious though.
Here we propose to express the partial
packing coefficient for each of the monomers via an integral of
the Gaussian trial RDF over the excluded volume area.
The resulting GSC equations involve well tractable,
albeit somewhat more complicated than before, expressions, since they now
depend on the exact shapes of the molecular potentials.

Although we shall present the resulting technique in its most general form,
which is applicable to any type of polymeric system, our numerical
analysis in this work will be restricted by homopolymers of three different
architectures --- ring, open and star --- across the coil--to--globule 
transition. 
This is done to enable us to make a comparison of the results 
from the present GSC
theory with the MC data obtained
in our previous paper in Ref. \onlinecite{CorFunc}.

\section{Model}
\label{sec:model}

The current coarse--grained polymer model is based
on the following Hamiltonian (energy functional) 
\cite{CorFunc,Torus,CopStar,CombStar} 
in terms of the monomer coordinates, ${\bf X}_i$:
\begin{eqnarray}
\frac{H}{k_B T} & = & \frac{1}{2\ell^2}  \sum_{i\sim j} \kappa_{ij}
        ({\bf X}_i - {\bf X}_{j})^2
      + \frac{1}{2\ell^2} \sum_{i\approx j \approx k} \lambda_{ijk}
        ({\bf X}_{i} + {\bf X}_{k} - 2{\bf X}_j)^2  \nonumber \\
  & & + \frac{1}{2} \sum_{ij,\ i\not= j}
\left( U^{(lj)}_{ij} (|{\bf X}_i - {\bf X}_j|)
+U^{(cou)}_{ij} (|{\bf X}_i - {\bf X}_j|)\right).
\label{cmc:hamil}
\end{eqnarray}
Here the first term represents the connectivity structure of the polymer with
harmonic springs of a given strength $\kappa_{ij}$, introduced between
any pair of connected monomers (which is denoted by $i\sim j$).
The second term represents the bending energy penalty given by the
square of the local curvature with a characteristic stiffness $\lambda_{ijk}$
between any three consecutively connected monomers (which is
denoted by $i\approx j \approx k$) in the form of the Kratky--Harris--Hearst.
Below we shall prefer to rewrite the first two terms in the following 
equivalent form,
\begin{equation}
\label{harmEq}
\frac{H^{(bond)}}{k_B T}=\frac{1}{2}\sum_{ij} U_{ij}^{(bond)}\, 
({\bf X}_i - {\bf X}_{j})^2. 
\end{equation}
Finally, the third  and fourth terms represent 
pair--wise non--bonded interactions
between monomers such as the van der Waals and Coulomb forces.
We can adopt the Lennard--Jones form of the former potential,
\begin{equation}\label{VLJ}
U^{(lj)}_{ij}(r) = \left\{
\begin{array}{ll}
+\infty, &  r < r^{(0)}_i + r^{(0)}_j\\
U_{ij}^{(0)} \left( \left( \frac{r^{(0)}_i + r^{(0)}_j}{r}\right)^{12}
- \left( \frac{r^{(0)}_i + r^{(0)}_j}{r} \right)^{6} \right), &
r > r^{(0)}_i + r^{(0)}_j
\end{array}
\right.,
\end{equation}
where there is also a hard core part with the monomer radii $r^{(0)}_i$,
and where $U_{ij}^{(0)}$ are the dimensionless strengths of the interactions. 
The Coulomb interaction potential similarly is,
\begin{equation}
\label{Coul}
U^{(cou)}_{ij}(r)=q_i\,q_j \frac{l_B}{r}\,\exp(-r/l_D),
\qquad l_B \equiv \frac{Q_0^2}{4\pi \epsilon_0 \,k_B T},
\end{equation}
where $l_D$ is the Debye screening length, $\epsilon_0$ is
the dielectric permittivity of vacuum, $Q_i=Q_0\,q_i$ are the
charges, and $l_B$ is the Bjerrum length.

\section{Method}
\label{sec:method}

\subsection{Equations of the GSC method}

The main objects in the GSC method are the mean--squared distances
between monomers,
\begin{equation} \label{eq:Ddef}
{\cal D}_{ij}(t) \equiv \frac{1}{3}\biggl\langle
({\bf X}_{i}(t) - {\bf X}_{j}(t))^2 \biggr\rangle.
\end{equation}
Note that our convention includes the factor of $1/3$ here and hence in the 
later definition of the mean--squared radius of gyration according
to the tradition. This allows us to rid of such factors from
the radial distribution function (RDF) and numerous averages over it. 

The GSC method is based on replacing the stochastic ensemble for
$\bbox{X}_i$ with the exact Hamiltonian in 
the Langevin equation of motion onto the trial ensemble 
$\bbox{X}_i^{(0)}(t)$ with a trial
Hamiltonian $H^{(0)}(t)$. The latter is taken as a generic quadratic form with
the matrix coefficients, which are called the time--dependent 
effective potentials,
\begin{equation} \label{gsc:trial}
H^{(0)} [{\bf X} (t)] = \frac{1}{2}\sum_{ij} V_{ij} (t)\,
{\bf X}_i (t)\, {\bf X}_{j} (t).
\end{equation}
Then one requires that the inter--monomer correlations satisfy the condition,
\begin{equation}
\langle \bbox{X}_{i}(t) \bbox{X}_{j}(t) \rangle_0 =
\langle \bbox{X}_{i}^{(0)}(t) \bbox{X}_{j}^{(0)}(t) \rangle_0,
\end{equation}
which means that the trial ensemble well approximates the exact one
as far as monomer correlations are concerned.
This procedure yields expressions for the effective potentials via the
instantaneous mean energy,
\begin{equation}
V_{ij} (t) = -\frac{2}{3}\frac{\partial {\cal E}[{\cal D}(t)]}
{\partial {\cal D}_{ij}(t)},
\label{gsc:effV}
\end{equation}
and hence the mean--squared distances themselves
satisfy the self--consistent equations \cite{CopStar}. These,
in the absence of the hydrodynamic interaction, are simply,
\begin{equation} \label{eq:kinmain}
\frac{\zeta_b}{2}\frac{d}{dt}{\cal D}_{ij}(t) = -\frac{2}{3}
\sum_{k}({\cal D}_{ik}(t)-{\cal D}_{jk}(t))\left(
\frac{\partial {\cal A}[{\cal D}(t)]}{\partial {\cal D}_{ik}(t)}-
\frac{\partial {\cal A}[{\cal D}(t)]}{\partial {\cal D}_{jk}(t)}
\right).
\end{equation}
Here $\zeta_b$ is the friction coefficient of a monomer,
and the instantaneous free energy has the same functional
expression via the instantaneous ${\cal D}_{ij}(t)$ as it
has at equilibrium.
Extension to the preaveraged hydrodynamic approximation is
quite straightforward also and it is discussed
in Ref. \onlinecite{ConfTra}.

The stationary limit of these equations produces the equations for
the minimum of the free energy, which are the same as
those derived from the Gibbs--Bogoliubov
variational principle with a generic quadratic trial Hamiltonian.
Although in this paper we shall only be concerned with the equilibrium
properties, the numerical solution of Eq. (\ref{eq:kinmain}),
applied until the stationary limit is reached,
presents by far the most efficient technique
for finding the global free energy minimum.
This, based on the fifth order adaptive step Runge--Kutta
integrator \cite{ConfTra}, was used for obtaining the results
from the GSC technique in this work.

It should be also noted that the systems studied here possess a large
number of kinematic symmetries for ${\cal D}_{ij}$, and hence for
$V_{ij}$, matrices coming from their symmetricity and
from the equivalence of any monomer in a ring,
or any arm in a star homopolymers. Thus, the computational expenses
per step in our calculations are of order $t_c \sim N\,F$, 
where $N/2 \lesssim F \lesssim N^2/2$ is the total number
of independent elements in the matrix ${\cal D}_{ij}$.
These symmetries significantly reduce the computational times
compared to MC for an equivalent system,
where such symmetries only appear in the observables
after averaging over the statistical ensemble. 
For comparison, the computational expenses per step in 
MC are of order $t_c \sim N\,\Delta t\,S$,
where $\Delta t\sim N^2$ is the number of MC steps
needed to ensure a good statistical independence
between measurements, and where $S$ is the number of measurements
needed for sampling of observables. Typical values of $S$
should be of order of $10^4-10^6$ for a good accuracy
in the present case \cite{CorFunc}.
Moreover, kinetic iteration of the GSC equations towards the 
equilibrium is also significantly faster than the equivalent
equilibration procedure in the MC.
For example, for a ring polymer of $N=150$ units in the good solvent,
$U^{(0)}=1$, the GSC method turns out to be $\approx 1000$ times
faster \cite{CPU} than MC for gaining the same data,
whereas for an open chain of the same length,
which possesses much fewer kinematic symmetries, GSC is
$\approx 200$ times faster than MC.

\subsection{Hard sphere contribution}

Carnahan and Starling have devised a simple but rather accurate
equation of state for hard sphere liquids \cite{CarnahanStarling}
in terms of the packing coefficient $\eta$, yielding
the following free energy,
\begin{equation}
\label{CS}
\frac{{\cal A}^{(CS)}}{k_B T\,N} = F^{(CS)}(\eta) =
\frac{\eta (4 - 3\eta)}{(1 - \eta)^2},
\qquad \eta = N\frac{v_0}{V},
\end{equation}
where $v_0$ and $V$ are the volumes of the hard sphere
and of the whole system respectively.
This is obtained from an interpolation formula for the virial coefficients
based on several of them known exactly (see more detailed
discussions in e.g. Ref. \onlinecite{HansenMacDonald}). 

To extend these ideas to polymeric fluids, we would have
to distinguish the packing coefficients for individual monomers. 
In Eq. (\ref{CS}) we can write $N=\sum_{i}$; the
volume of the sphere $v_0$ is equal to the $1/8$-th of its excluded volume; and
the inverse volume is equal to the RDF of 
the ideal reference system, $1/V=g^{(2)}_{ideal}$. Thus, we can similarly
express the packing coefficient for the $i$-th monomer
as a sum over all other monomers of the $1/8$-th of the integral over
the excluded volume of the monomer $i$ of the 
RDF for the Gaussian reference system,
\begin{equation}
\label{gGaus}
g^{(2)}_{ij}({\bf r})=\frac{1}{(2\pi\,{\cal D}_{ij})^{3/2}}\,
\exp\left( -\frac{{\bf r}^2}{2{\cal D}_{ij}}\right),
\end{equation}
yielding finally,
\begin{equation}
\label{eta}
\eta =N\frac{v_0}{V} \quad \longrightarrow \quad \eta_i = 
\sum_{i\neq j}\,\frac{a}{8}
\int_{|{\bf r}| < r^{(0)}_i+r^{(0)}_j} d{\bf r}\, g^{(2)}_{ij}({\bf r}).
\end{equation}
One may note that the Gaussian distribution does not possess
a well defined volume beyond which it vanishes.
Thus, to account for this we can, 
in principle, include a multiplicative parameter $a$,
which should be once and for all chosen to match the data best, 
but this should, in any case, be fairly close to the unity.
Therefore, the total hard sphere contribution will be,
\begin{equation}
\label{Shs}
\frac{{\cal A}^{(hs)}}{k_B T}= N\,F^{(CS)}(\eta) \quad
\longrightarrow \quad \sum_i F^{(CS)}(\eta_i).
\end{equation}

\subsection{Free energy in the GSC method}

The total mean energy includes both the bonded and the
pair--wise non--bonded interactions via,
\begin{equation}
\label{Energy}
\frac{{\cal E}^{(int)}}{k_B T}
=\frac{1}{2} \sum_{ij} \left( 3 U^{(bond)}_{ij}\,{\cal D}_{ij}
+\int_{r>r_i^{(0)}+r_j^{(0)}}d{\bf r}\,g^{(2)}_{ij}({\bf r})
\,[U^{(lj)}_{ij}({\bf r})+U^{(cou)}_{ij}({\bf r})] \right).
\end{equation}
Note that here we integrate only beyond the excluded volume as the hard sphere
contribution will be included explicitly via the Carnahan--Starling term.

Given that the conformational Gaussian entropy has been 
calculated by us in Appendix B of Ref. \onlinecite{ConfTra},
the total free energy can be summarised as follows,
\begin{equation}
{\cal A} \equiv {\cal E} - T {\cal S}, \quad
{\cal E}= {\cal E}^{(int)} + {\cal A}^{(hs)}, \quad {\cal E}^{(int)}= 
{\cal E}^{(bond)} + {\cal E}^{(lj)} + {\cal E}^{(cou)}, \quad 
{\cal S} \equiv {\cal S}^{(gau)},
\end{equation}
where its various terms are given by,
\begin{eqnarray}
\frac{{\cal E}^{(bond)}}{k_B T} & = & 3 \sum_{ i< j} U^{(bond)}_{ij}
                        {\cal D}_{ij}, \\
\frac{{\cal E}^{(lj)}}{k_B T}   & = & \sum_{i<j} U_{ij}^{(0)}
                        E^{(lj)} [ y_{ij} ], \\
\frac{{\cal E}^{(cou)}}{k_B T}  & = &  \sum_{i < j} 
\frac{l_B\, q_i q_{j}}{r^{(0)}_i+r^{(0)}_j}
                        E^{(cou)} [ y_{ij}, (r_i^{(0)}+r_{j}^{(0)})/l_D ], \\
\frac{{\cal S}^{(gau)}}{k_B} & = & \frac{3}{2} \ln {\rm det}\, R^{(N-1)}, \\
\frac{{\cal A}^{(hs)}}{k_B T}  & = & \sum_i F^{(CS)}(\eta_i), \qquad
F^{(CS)}(\eta_i)=\frac{\eta_i (4 - 3\eta_i)}{(1 - \eta_i)^2},
\end{eqnarray}
and the arguments of these functions are defined as,
\begin{eqnarray}
\eta_i & = & a\sum_{j \neq i} F^{(\eta)} [ y_{ij} ], \label{eq:etai} \\
y_{ij} & = & \frac{\sqrt{{\cal D}_{ij}}}{r_i^{(0)} + r_{j}^{(0)}}, 
\label{eq:Y}
\\
R_{ij} & = & \frac{{\cal D}_i + {\cal D}_{j}}{2} - {\cal R}_g^2 - 
\frac{{\cal D}_{ij}}{2},
\quad {\cal D}_i \equiv \frac{1}{N}\sum_{j}{\cal D}_{ij}, 
\quad {\cal R}_g^2 \equiv \frac{1}{2N^2}\sum_{ij}{\cal D}_{ij}.
\end{eqnarray}
Finally, the functions of $y_{ij}$ are,
\begin{eqnarray}
F^{(\eta)} [y] & \equiv & \frac{1}{(2y)^3} \sqrt{\frac{2}{\pi}}
                 \int_0^1 x^2\, dx\, \exp\left( -\frac{x^2}{2 y^2} \right)
                 = \nonumber\\
         & = & \frac{y\,{\rm erf}\left(\frac{1}{y\sqrt{2}}\right) -
                \sqrt{2/\pi}\,\exp\left(-\frac{1}{2y^2}\right)}{8y},
\label{eq:Fy} \\
E^{(lj)}   [y] & \equiv & \frac{1}{y^3} \sqrt{\frac{2}{\pi}}
                 \int_1^{\infty} x^2\, dx\,
                 \exp\left( -\frac{x^2}{2 y^2} \right)\,
                 \left( \frac{1}{x^{12}} - \frac{1}{x^6} \right)
                 = \nonumber\\
         & = & \frac{ \sqrt{2/\pi}\,\exp\left(-\frac{1}{2y^2}\right)
               \left(y-y^3+3y^5+300y^7-210y^9\right) -
               {\rm erfc}\left(\frac{1}{y\sqrt{2}}\right)
               \left(1+315y^6\right)}{945 y^{12}},\\
E^{(cou)}  [y, k] & \equiv & \frac{1}{y^3} \sqrt{\frac{2}{\pi}}
               \int_1^{\infty} x^2\,dx\, \exp\left(-\frac{x^2}{2 y^2}\right)\,
               \frac{\exp(-k x)}{x}
               = \nonumber\\
         & = & \exp\left(-\frac{1}{2y^2} - k\right)
               \left( \frac{1}{y}\sqrt{\frac{2}{\pi}} - \frac{2k}{\pi}
               \exp\left(\frac{(1+ky^2)^2}{2y^2}\right)\,
               {\rm erfc}\left(\frac{1+ky^2}{y\sqrt{2}}\right)\right).
\end{eqnarray}
Here we have used the standard definitions for the error functions,
\begin{equation}
\label{erf}
{\rm erf}(z)\equiv \frac{2}{\sqrt{\pi}}\int_0^z dx\, \exp(-x^2),
\qquad {\rm erfc}(z) \equiv 1 - {\rm erf}(z).
\end{equation}
Note also that for a large $z$, numerically, one has to use
the truncated asymptotic expansion in calculating $E^{(cou)}$
to avoid divergences, 
\begin{equation}
\label{appr}
\mbox{erfc}(z)\exp(z^2) \approx \frac{1}{z\,\sqrt{\pi}}\left(
1-\frac{1}{2z^2}+\frac{3}{4z^4}-\frac{15}{8z^6}+\ldots
\right).
\end{equation}

\section{Results}\label{sec:results}

Here we shall restrict ourselves by the case of homopolymers, 
so that in Eq. (\ref{cmc:hamil}) all non--zero bonded interaction
constants are equal: $\kappa_{ij}=\kappa$, $\lambda_{ijk}=\lambda$, 
as well as all non--bonded interaction parameters are identical 
in Eq. (\ref{VLJ}): $U^{(0)}_{ij}=U^{(0)}$, $r^{(0)}_i\equiv d/2$.
We also choose the hard sphere diameter $d$
equal to the length $\ell$  defined in Eq. (\ref{cmc:hamil}),
as in Ref. \onlinecite{CorFunc}.
Moreover, henceforth we shall use the mean energy ${\cal E}^{(int)}$
expressed in units of $k_B T$ and the mean--squared
distances ${\cal D}_{ij}$ and the mean--squared 
radius of gyration $3 {\cal R}_g^2$ expressed
in units of $\ell^2$.

Firstly, to understand the influence of the adopted Carnahan--Starling
term, we shall look at the case of a ring homopolymer with varied
values of the spring constant in the good athermal solvent, $U^{(0)}=0$. 
In Tab. \ref{tab:1} we compare values
of the mean energy ${\cal E}^{(int)}$ between MC and GSC
with two choices of the multiplicative parameter
$a$ in Eq. (\ref{eq:etai}) for the packing coefficient,
namely $a=0.9$ and $a=1.0$. Likewise, in Tab. \ref{tab:2} we present
the data for the mean--squared radius of gyration $3{\cal R}_g^2$.
One can see that the results from the GSC theory simply coincide with those
from MC as $\kappa \to 0$. Yet, the agreement in the energy is somewhat better
for the theory with $a=0.9$ than for that with the naive choice $a=1.0$, and
that is how exactly the particular value $a=0.9$ has been chosen by us.
A typical relative error in the energy is less than $1.5$ percent for
$a=0.9$ and is under $4$ percent for $a=1.0$. The agreement for
the mean--squared radius of gyration is somewhat less impressive ---
the relative error steadily increases with $\kappa$, being better
for $a=0.9$, but never exceeding a few dozen percent.

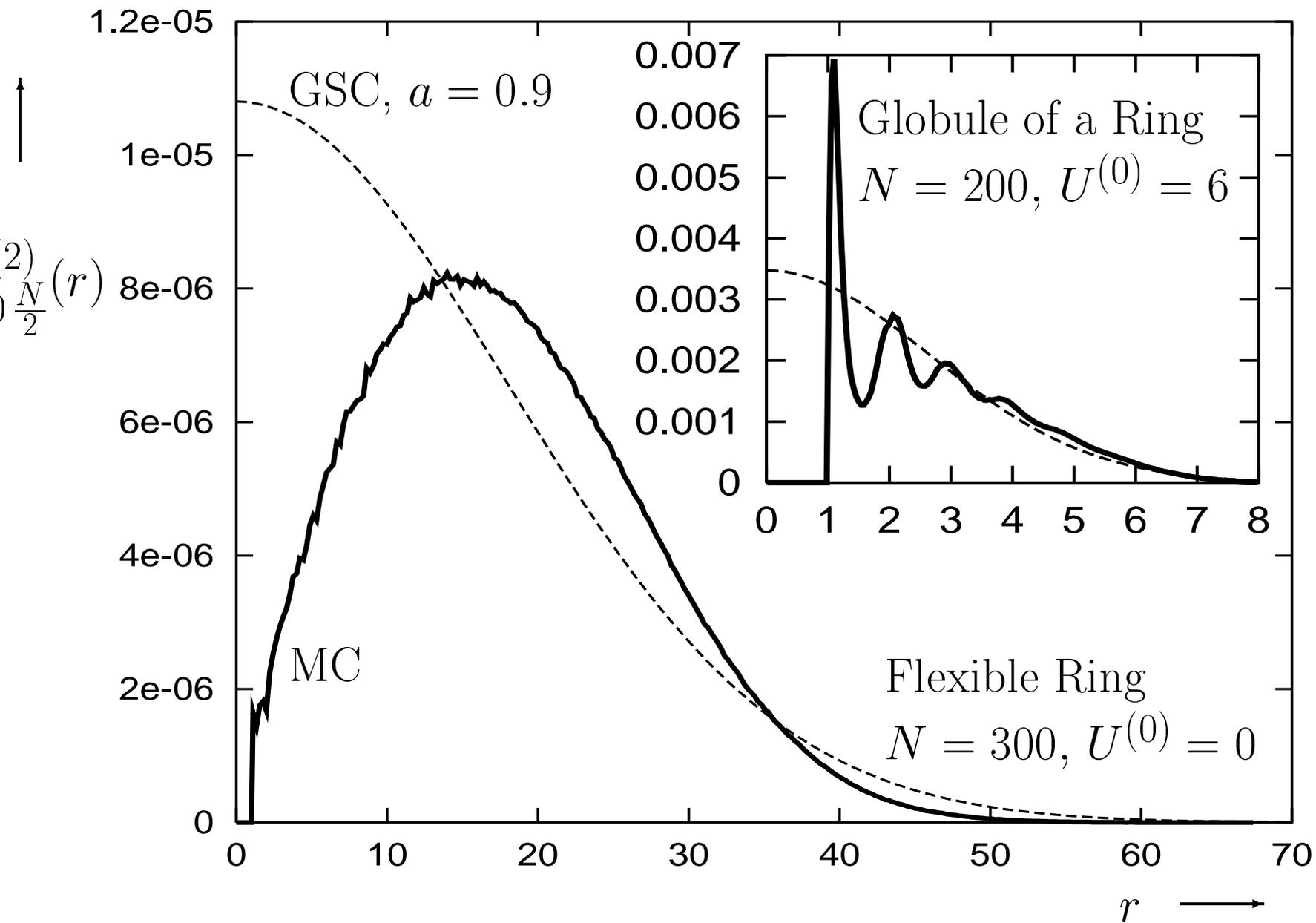
\begin{figure}
\caption{ \label{fig:RDF}
Plots of the half--ring radial distribution functions 
$g^{(2)}_{0\,N/2}(r)$ (in $\ell^{-3}$ units) from MC
(solid thick lines) and the GSC theory with $a=0.9$
(thin dashed lines) vs the radial separation $r$ (in $\ell$ units)
for the flexible ring homopolymers with $\lambda=0$ and $\kappa=1$. 
The main part of the figure corresponds to the good
solvent $U^{(0)}=0$ and the polymer size $N=300$, whereas
the inset corresponds to the poor solvent $U^{(0)}=6$ and 
the polymer size $N=200$.
}
\end{figure}

One can comment on the reason why a value $a<1$ produces a somewhat
better agreement with MC.
To facilitate this discussion, RDFs from the GSC and MC techniques are exhibited
in Fig.~\ref{fig:RDF}.
In Ref. \onlinecite{CorFunc} we have discussed the role of
the `correlation hole' at small separations beyond the
excluded volume area, which is also evident
for $r\lesssim 15$ in the main part of Fig.~\ref{fig:RDF}.
As this feature
is absent in the Gaussian theory (see Eq. (\ref{gGaus})),
the resulting packing coefficient values in Eq. (\ref{eq:etai})
are overestimated. Thus, reducing the parameter $a$ permits us to effectively
lower the packing coefficient and this can be done only once
because the correlation hole effect is expressed in terms
of the function of the dimensionless variables, 
\begin{equation}
\hat{g}^{(2)}_{ij}(\hat{r}) \equiv {\cal D}_{ij}^{3/2}\,\,
g^{(2)}_{ij}\left(r\right),
\qquad \hat{r} \equiv \frac{r}{{\cal D}_{ij}^{1/2}}.
\end{equation}

Next, we would like to analyse the dependence on the degree of
polymerisation, $N$, in the good solvent, $U^{(0)}=0$,
(the second block in Tabs.~\ref{tab:1}, \ref{tab:2}).
The mean energy in this case comes entirely from the bonded
interactions and hence the relative energy deviation of GSC from MC
does not increase with $N$. This means simply that the values of
${\cal D}_{i,i+1}$ match with those from MC quite well. As for the radius of
gyration, the disagreement increases slowly with $N$, i.e. the
mean--squared distances ${\cal D}_{i,i+k}$ are
increasingly more overestimated by GSC as compared to MC for large $k$.
Based on these results, we can determine the swelling exponent
of the ring coil by fitting $3{\cal R}_g^2$ via Eq.,
\begin{equation}
\label{Rg2Nu}
3{\cal R}_g^2=b^2\,N^{2\nu}
\end{equation}
using our data in the range $N=50-500$. The resulting prefactor $b^2$
and the exponent $\nu$ in all three cases are,
\begin{eqnarray}
b^2 & = & \left\{ 
\begin{array}{ll}
0.223\pm 0.005 & \quad {\rm MC} \\
0.180\pm 0.003 & \quad {\rm GSC},\ a=0.9 \\
0.184\pm 0.003 & \quad {\rm GSC},\ a=1.0
\end{array}
\right.   \label{BFit}\\
\nu & = & \left\{ 
\begin{array}{ll}
0.610 \pm 0.006 & \quad {\rm MC} \\
0.648 \pm 0.002 & \quad {\rm GSC},\ a=0.9 \\
0.651 \pm 0.001 & \quad {\rm GSC},\ a=1.0
\end{array}
\right.   \label{NuFit}
\end{eqnarray}
Thus, the increase in the deviation of $3{\cal R}_g^2$ in the GSC method from MC
indeed comes entirely from the overestimation
of the swelling exponent. This is believed to reach the value
$\nu=2/3=0.666\ldots$ asymptotically.
However, in the present range of $N$ both GSC and
MC give higher apparent exponent values than the most accurate
renormalisation group result up--to--date \cite{Guida}, 
$\nu=0.5882 \pm 0.0011$.
The GSC results for $\nu$ are only slightly sensitive on the value of $a$,
being in a relatively small overestimation over the MC results.
This is related to the Gaussian shape of $g^{(2)}_{ij}$ in the GSC theory,
whereas a stretched exponential tail, $\exp(-B\hat{r}^\delta)$, 
(see the range $r\gtrsim 40$ in Fig.~\ref{fig:RDF})
contributes most to $3{\cal R}_g^2$ in MC.
We may note also that the GSC estimate for $\nu$ is fairly close
to the result $\nu=0.635$ from the integral equations approach
based on a complicated closure of the Born--Green-Yvon hierarchy
in Ref. \onlinecite{Taylor}.

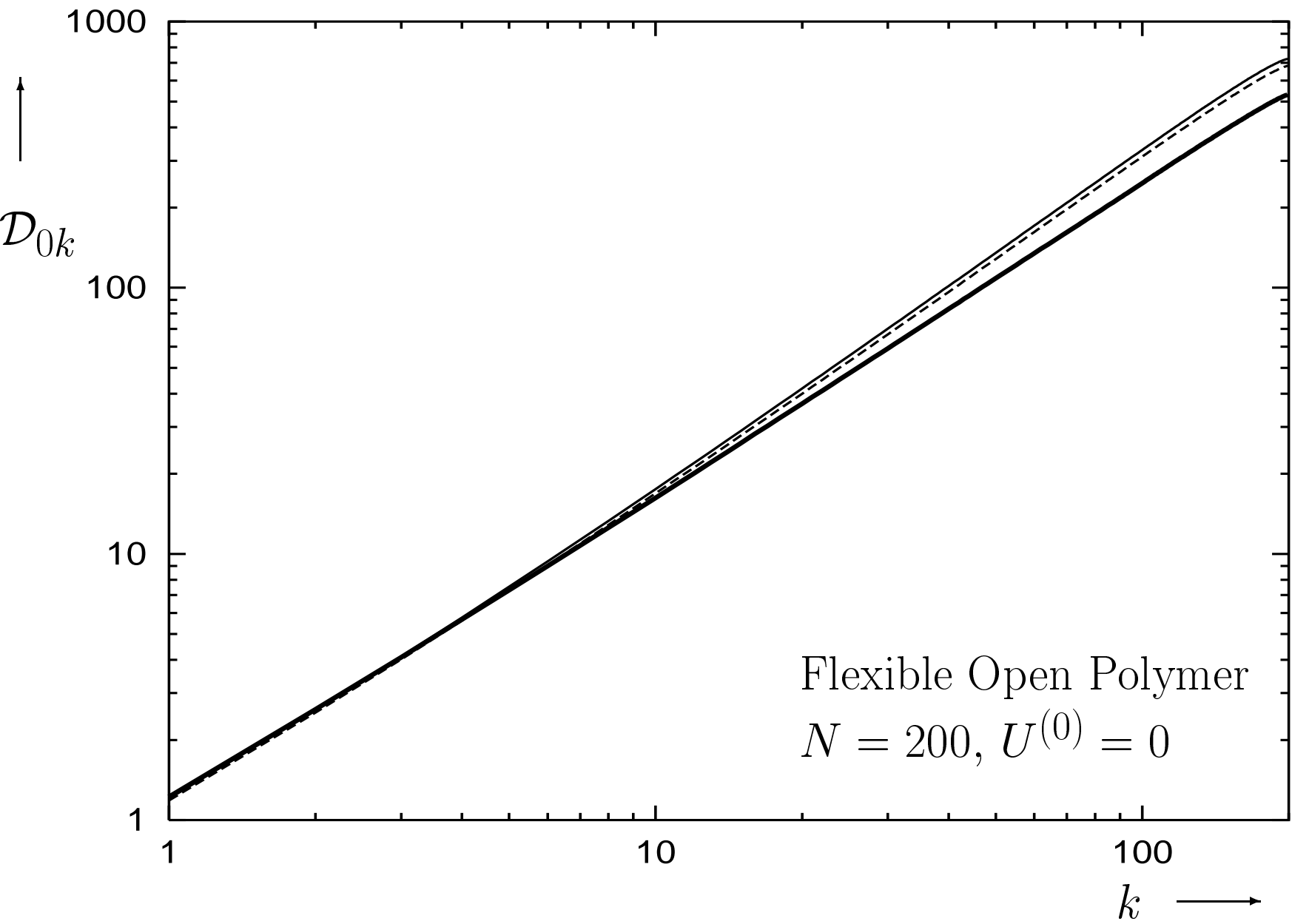
\begin{figure}
\caption{ \label{fig:DkOpen}
The mean--squared distances ${\cal D}_{0k}$ (in $\ell^2$ units)
of an open flexible homopolymer with $N=200$, $\lambda=0$, and $\kappa=1$ 
in the good solvent, $U^{(0)}=0$, vs the chain index $k$.
Here and below the solid thick lines correspond to the MC data,
the solid thin lines to the GSC theory with $a=1.0$,
and the dashed lines to the GSC theory with $a=0.9$.
}
\end{figure}

When considering open flexible homopolymers ($\kappa=1$, $\lambda=0$)
in the good solvent ($U^{(0)}=0$), the general behaviour of ${\cal E}^{(int)}$ 
and $3{\cal R}_g^2$ in the third block of Tabs. \ref{tab:1}, \ref{tab:2}
is very similar to that of a ring: a good and almost
$N$-independent agreement for ${\cal E}^{(int)}$ and a slow increase with $N$
in the relative error for $3{\cal R}_g^2$. In Fig.~\ref{fig:DkOpen}
we plot the mean--squared distances from the end monomer ${\cal D}_{0k}$
vs the chain index $k$. Up to $k \sim 10$, the agreement of both
GSC curves with the MC data is nearly perfect, whereas both GSC curves 
increasingly overestimate the MC data for larger $k$. This is consistent
with the overestimation of $3{\cal R}_g^2$ by the GSC method, dominated
by large $k$ contributions. Notably, the effect of changing the parameter
$a$ is rather weak on this scale.

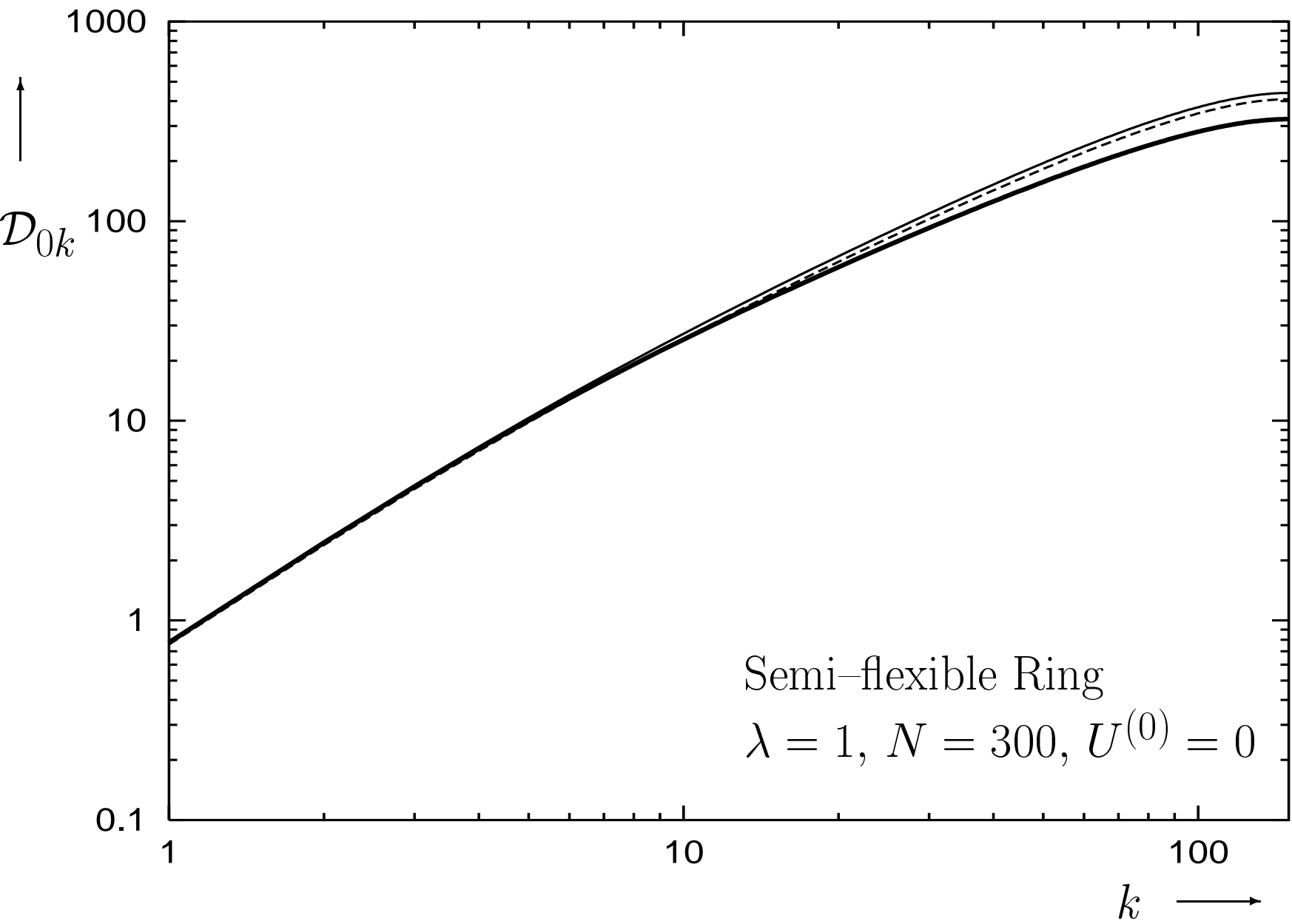
\begin{figure}
\caption{ \label{fig:DkRing}
The mean--squared distances ${\cal D}_{0k}$ (in $\ell^2$ units)
of a fairly flexible, $\lambda=1$, ring homopolymer with $N=300$
and $\kappa=1$ 
in the good solvent, $U^{(0)}=0$, vs the chain index $k$.
}
\end{figure}

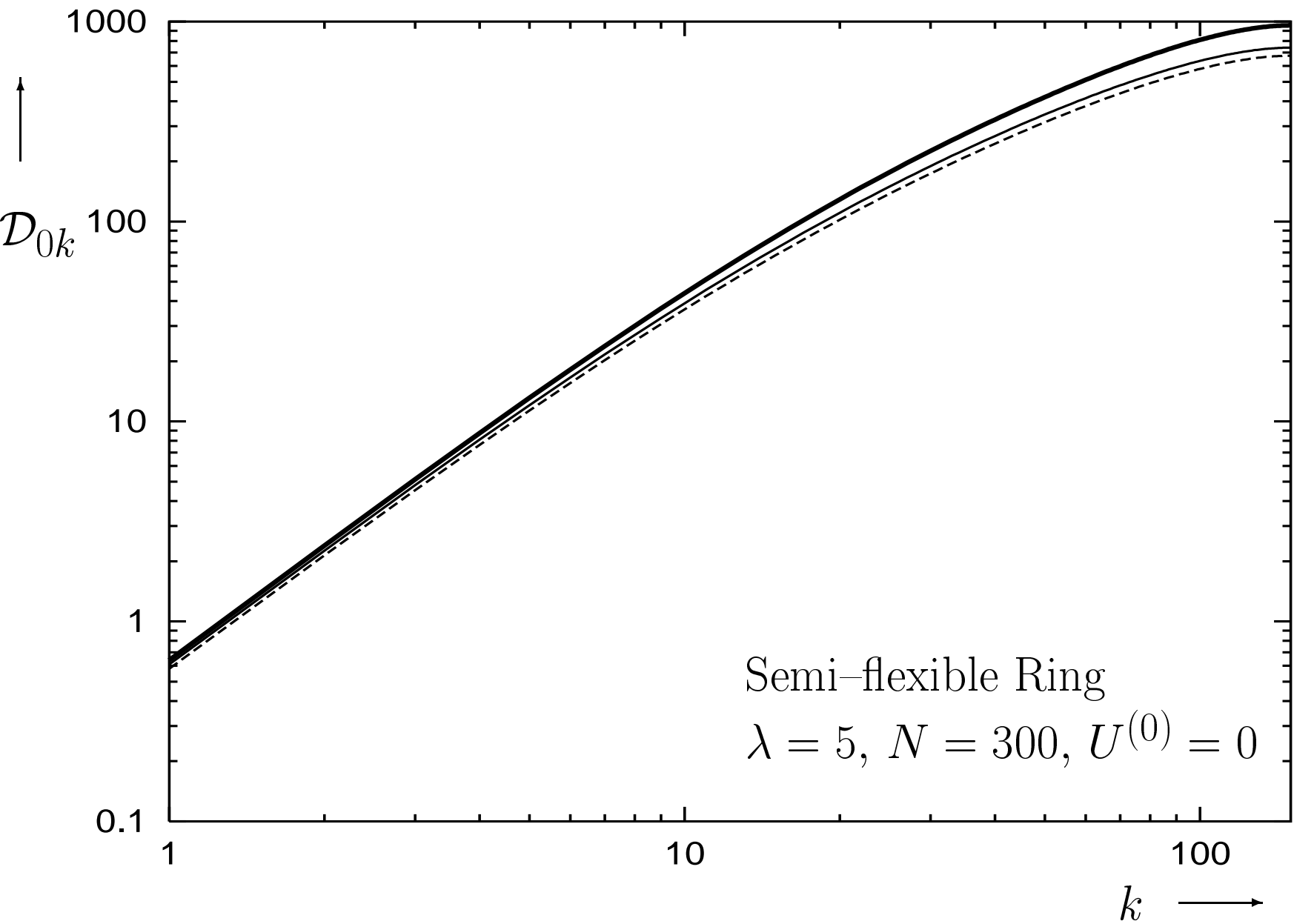
\begin{figure}
\caption{ \label{fig:DkRingStiff}
The mean--squared distances ${\cal D}_{0k}$ (in $\ell^2$ units)
of a semi--flexible, $\lambda=5$, ring  homopolymer with $N=300$
and $\kappa=1$ in the good solvent, $U^{(0)}=0$, vs the chain index $k$.
}
\end{figure}

Further, let us investigate the effect of increasing the chain stiffness
$\lambda$ from a value corresponding to a fairly flexible 
ring $\lambda=1$ (see Fig. \ref{fig:DkRing}
and the fourth block in Tabs. \ref{tab:1}, \ref{tab:2}) to that of a semi--stiff 
ring $\lambda=5$ (see Fig. \ref{fig:DkRingStiff}
and the fifth block in Tabs. \ref{tab:1}, \ref{tab:2}).
The fairly flexible case gives the energies in a very good agreement 
with the MC data, albeit the theory with $a=0.9$ is somewhat less accurate 
accidentally. However, the agreement of $3{\cal R}_g^2$ in
the GSC theory with MC is even better here than for
the corresponding flexible coil.
The semi--flexible case also gives the energies in a good agreement 
with the MC data, while $3{\cal R}_g^2$ tends to be underestimated by
the GSC method.
Plots of the mean--squared distances ${\cal D}_{0k}$ vs the chain index
$k$ in Figs.~\ref{fig:DkRing}, \ref{fig:DkRingStiff} match nearly
perfectly up to $k\sim 10$ as well, diverging
for larger $k$. The MC curve is lower (higher) than the GSC
curves for $\lambda=1$ ($\lambda=5$) in accord with the tables data.
Thus, overall, the increase of $3{\cal R}_g^2$ with the chain stiffness 
$\lambda$ is more rapid in MC than in the GSC theory.
Indeed, conformations of a stiff chain in MC become those
of a rigid ring (or rod for an open chain) with increasing $\lambda$.
However, the Carnahan--Starling equation was deduced in the assumption of
a total 3-d isotropy, when its influence would be weaker than
in an effective 1-d projection.

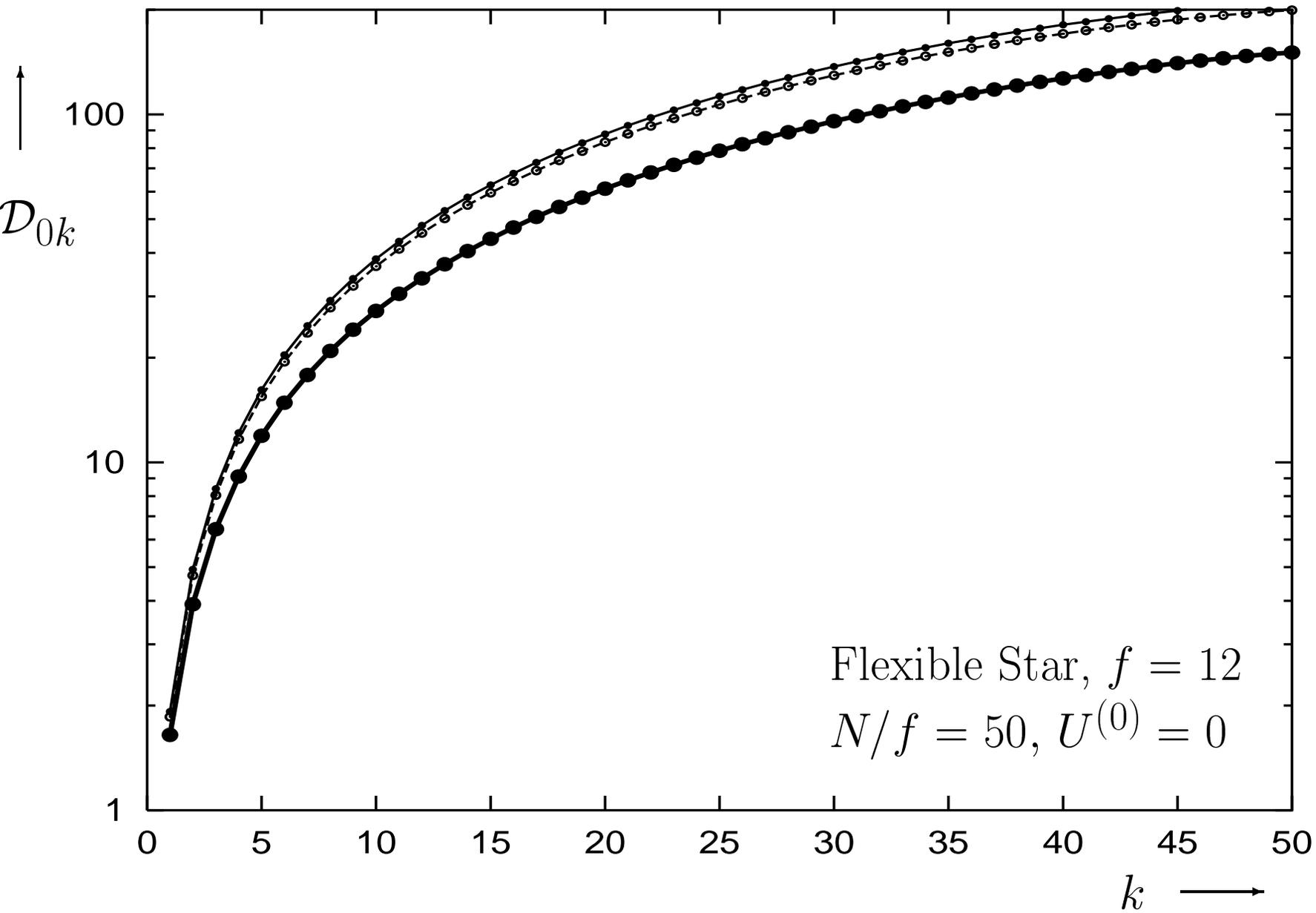
\begin{figure}
\caption{ \label{fig:DkStar}
The mean--squared distances ${\cal D}_{0k}$ (in $\ell^2$ units)
from the core monomer of a flexible, $\lambda=0$, homopolymer star
with $f=12$ arms, $\kappa=1$, and the arm length $N/f=50$ 
in the good solvent, $U^{(0)}=0$, vs the chain index $k$.
}
\end{figure}

Next, let us bring our attention to the effect of changing the topology
of the chain. Thus, in the sixth block of Tabs.~\ref{tab:1},
\ref{tab:2} we present ${\cal E}^{(int)}$ and $3{\cal R}_g^2$
for the flexible ($\lambda=0$) stars with the arm
length $N/f=50$ in the good solvent ($U^{(0)}=0$). 
The relative errors in ${\cal E}^{(int)}$ and $3{\cal R}_g^2$ increase
with the number of arms $f$ steadily, again the energy values being more
close between MC and GSC.
The mean--squared distances $D_{0k}$ 
from the core monomer for the largest star with $f=12$ arms 
are plotted in Fig.~\ref{fig:DkStar}. Clearly, the agreement
between GSC and MC is the worst of all previously
considered cases here. Even the values of ${\cal D}_{0k}$ 
do not match  for small $k$ because the core monomer is strongly affected
by the very pronounced correlation hole effect in MC \cite{CorFunc}.
However, ${\cal D}_{ij}$ between monomers within same arms
and away from the core are naturally closer between GSC and MC,
just as for the open chain in Fig.~\ref{fig:DkOpen}.
Clearly, the divergence of the GSC and MC curves does not
increase with $k$ after $k\sim 10$ and the curves have rather
similar overall shapes.

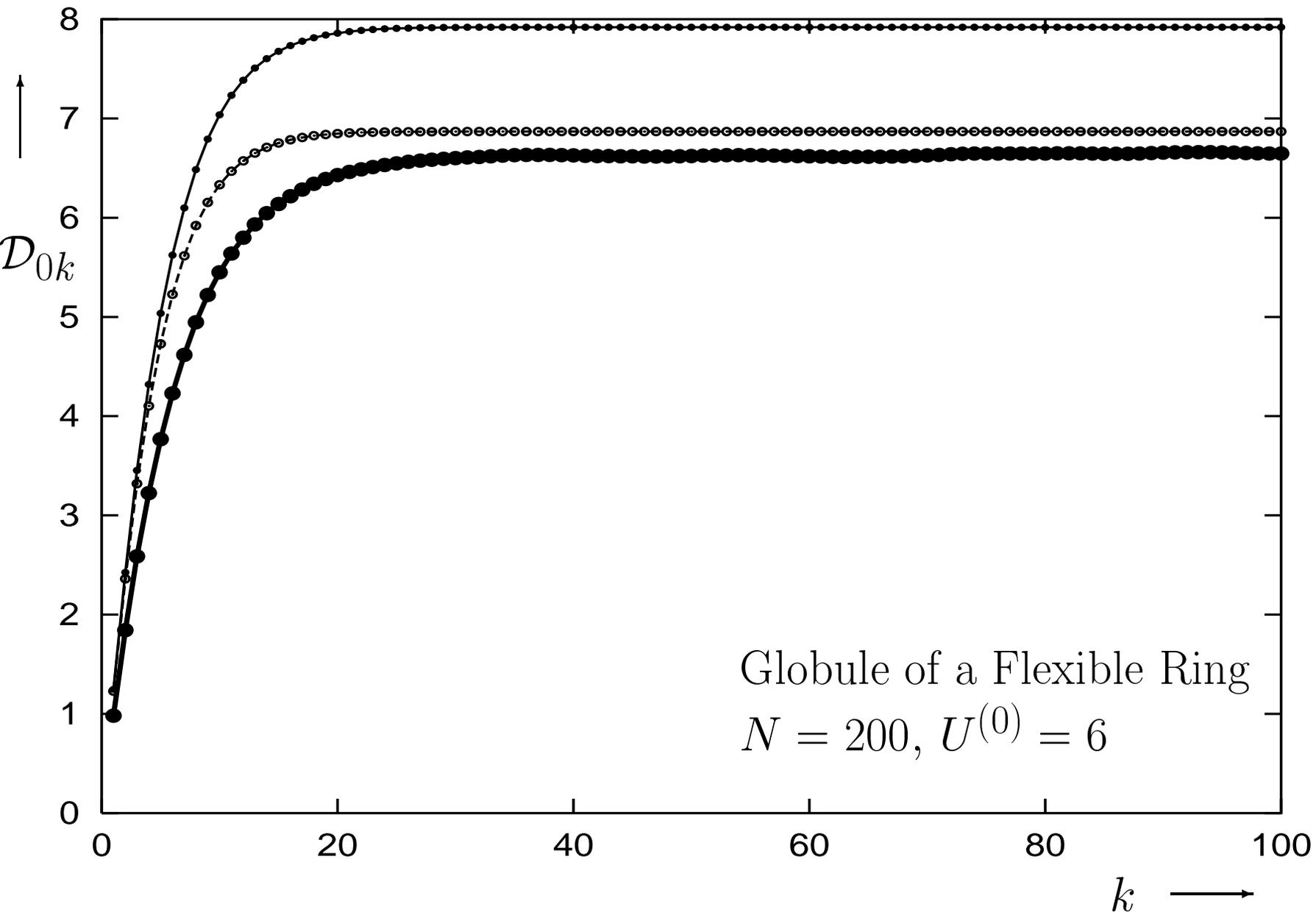
\begin{figure}
\caption{ \label{fig:DkGlob}
The mean--squared distances ${\cal D}_{0k}$ (in $\ell^2$ units)
for the globule of 
a flexible, $\lambda=0$, ring homopolymer 
with $N=200$, $\kappa=1$, and $U^{(0)}=6$ vs the chain index $k$.
}
\end{figure}

It is interesting to analyse the structure of the collapsed globule now.
Thus, in Fig.~\ref{fig:DkGlob} the mean--squared distances
${\cal D}_{0k}$ are plotted vs $k$ for a flexible ring
homopolymer in the globular state, $U^{(0)}=6$.
Overall shapes of the GSC and MC curves are quite similar, reflecting
the compactness of the globule. The discrepancy between 
GSC and MC at small $k$ is present systematically. GSC generally
overestimates the values of all ${\cal D}_{0k}$, but the theory with
$a=0.9$ manages to come to a nearly correct limit of the globule size 
for $k^*\gtrsim 20$ (see Ref. \onlinecite{CorFunc}), whereas 
the theory with $a=1.0$ is less accurate. 
The data for $3{\cal R}_g^2$ in the last block of Tab.~\ref{tab:2}
thus show a much better agreement than before, but the
energy values in Tab.~\ref{tab:1} have a significantly larger discrepancy
between GSC and MC. The GSC theory noticeably underestimates the
negative Lennard--Jones energy contribution between all pairs of
monomers. This, however, is to be expected given that the shape
of RDF from MC \cite{CorFunc}  
has a very tall first liquid--like peak (see the inset of Fig.~\ref{fig:RDF}).
Since the Lennard--Jones interaction is rather short--ranged,
the first peak gives a predominant negative contribution
to the mean energy in Eq. \ref{Energy}.
As the GSC theory has merely an effective smooth `interpolating' 
Gaussoid in $g^{(2)}_{ij}$ in that area
(see the inset of Fig.~\ref{fig:RDF}), the resulting negative 
energy contribution is significantly smaller in such a theory.

We can also remark that the scaling for the swelling exponent in the globule
is correct in the GSC theory. Indeed, the maximal compression is reached
when $\eta \lesssim 1$, therefore by considering $y_{ij}\to\infty$ in Eqs. 
(\ref{eq:etai},\ref{eq:Y},\ref{eq:Fy}) we indeed obtain, 
$D_{ij} \sim (r^{(0)}_i+r^{(0)}_j)^2\,N^{2/3}$.

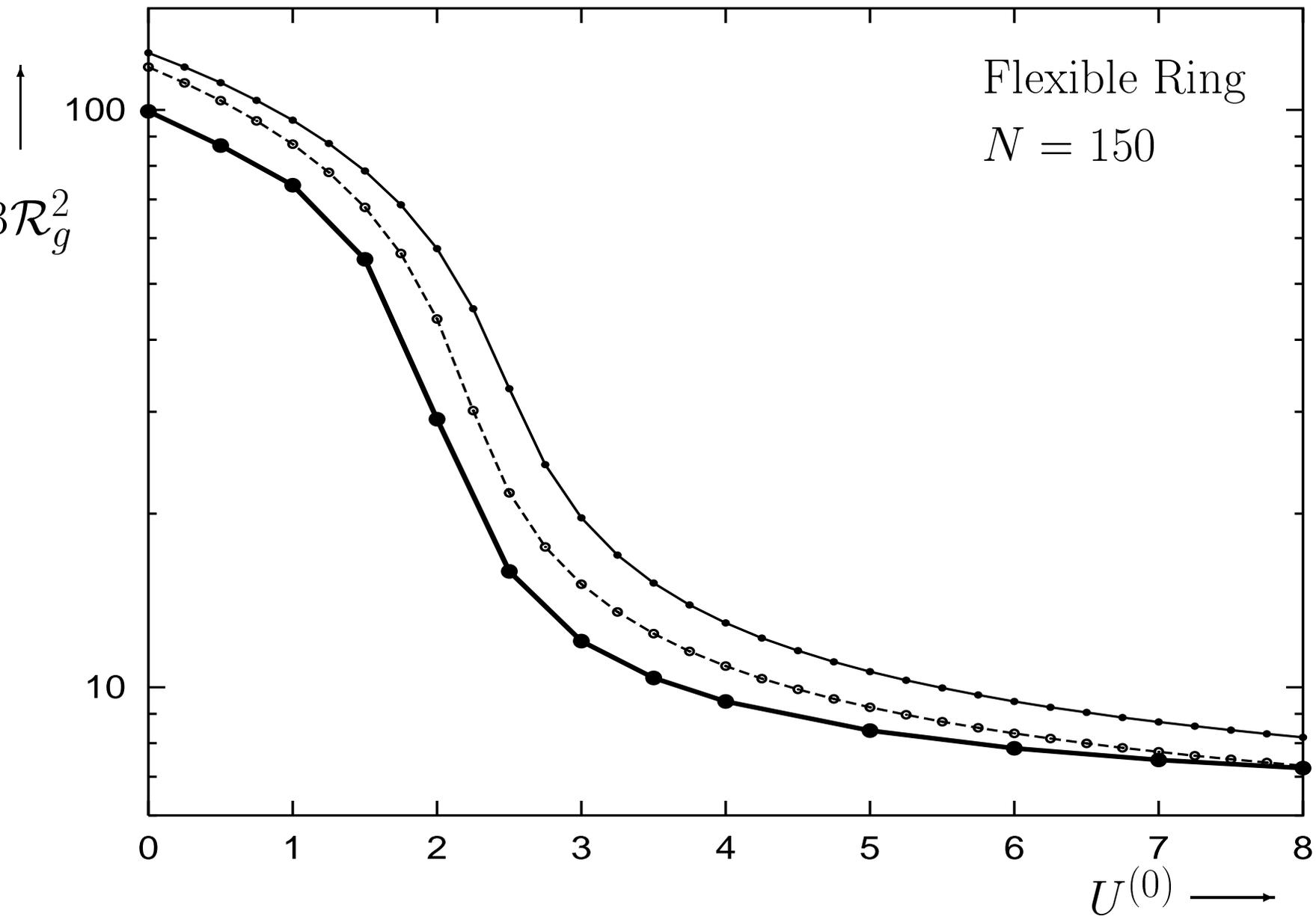
\begin{figure}
\caption{ \label{fig:RgCollapse}
The mean--squared radius of gyration $3{\cal R}_g^2$ (in $\ell^2$ units)
of a flexible, $\lambda=0$, homopolymer ring with $N=150$ and $\kappa=1$
vs the dimensionless degree of the Lennard--Jones attraction, 
$U^{(0)}$, across the coil--to--globule transition.
}
\end{figure}

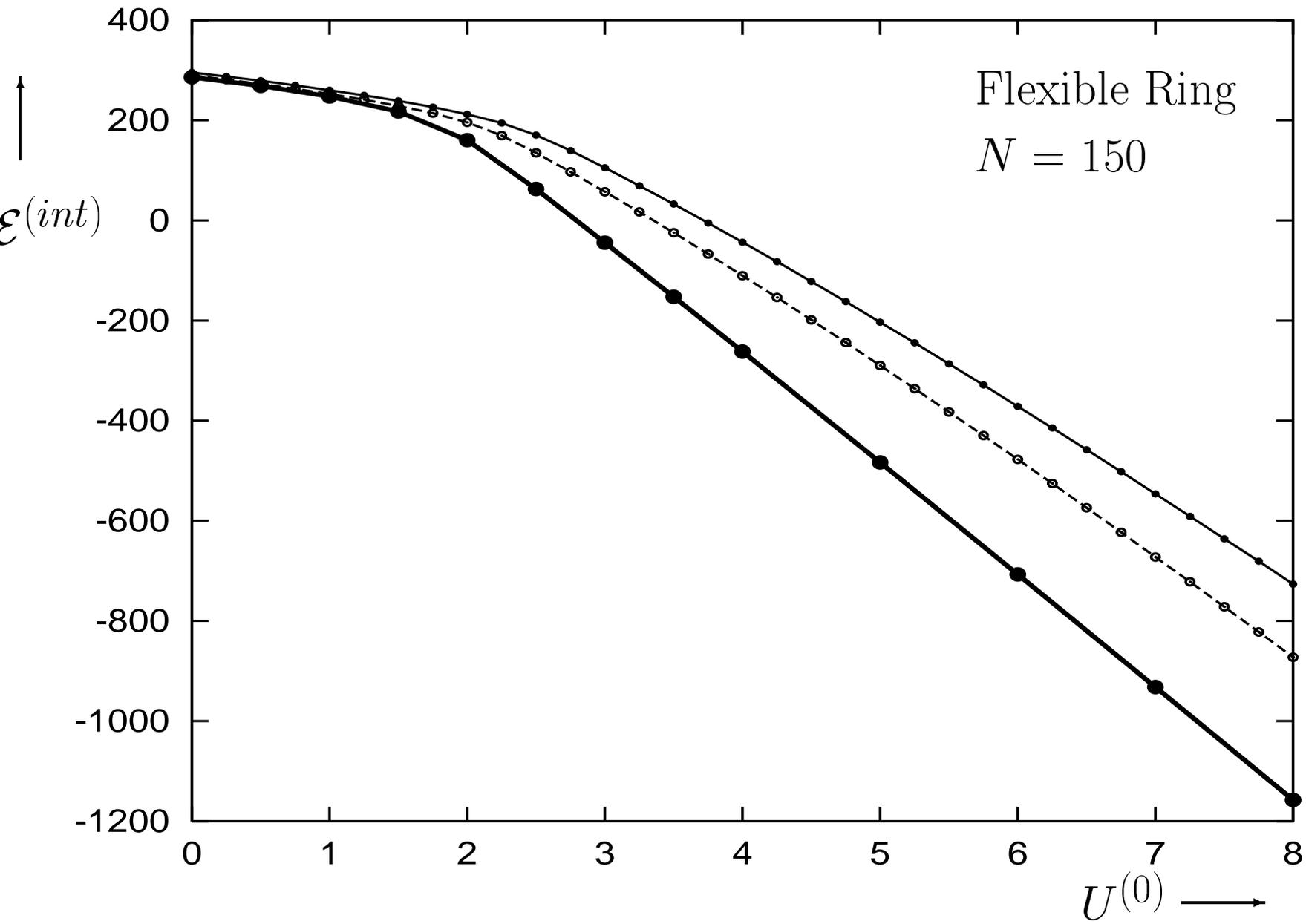
\begin{figure}
\caption{ \label{fig:ECollapse}
The mean energy ${\cal E}^{(int)}$ (in $k_B T$ units)
of a flexible, $\lambda=0$, homopolymer ring with $N=150$ and $\kappa=1$
vs the degree of the Lennard--Jones attraction, 
$U^{(0)}$, across the coil--to--globule transition.
}
\end{figure}

Next, we would like to look at the plots of the mean--squared
radius of gyration, $3{\cal R}_g^2$, and of the mean energy,
${\cal E}^{(int)}$ across the coil--to--globule transition. 
These are depicted in 
Figs.~\ref{fig:RgCollapse} and \ref{fig:ECollapse} respectively.
The following points can be made. First of all, the shapes of these
curves are quite similar for MC and GSC with both values of $a$.
Secondly, the coil--to--globule transition is continuous in all three cases,
with the energy slope changing noticeably at around the theta--point.
Thirdly, the transition occurs at a somewhat higher 
value of the attraction constant $U^{(0)}$ in the GSC theory
than in the MC simulation.
This can be explained by the underestimation of the Lennard--Jones
attraction energy in the globule discussed above.
Lastly, $3{\cal R}_g^2$ of the globule is partly overestimated
by the GSC method with $a=1$,
since the value $U^{(0)}=6$
is much closer to the point of the coil--to--globule transition 
for the theory with $a=1$ than that with $a=0.9$ or 
MC.

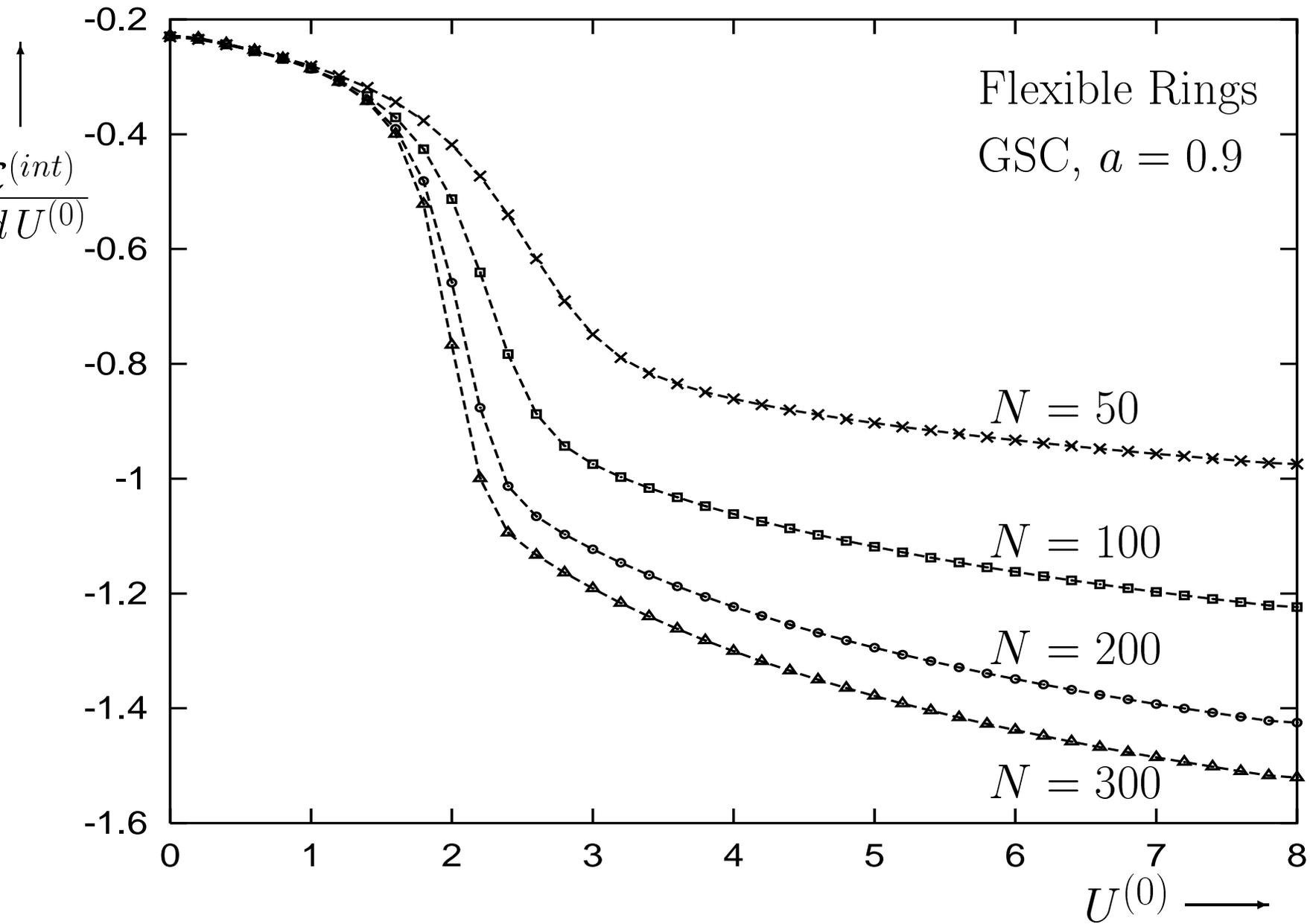
\begin{figure}
\caption{ \label{fig:CAP}
Plots of the specific energy slope, $N^{-1}d{\cal E}^{(int)}/dU^{(0)}$,
(in $k_B T$ units) of flexible, $\lambda=0$, homopolymer rings with
$\kappa=1$ vs the degree of the Lennard--Jones attraction, 
$U^{(0)}$, across the coil--to--globule transition
for different polymer sizes $N=50,100,200,300$ (from top to bottom).
These are obtained from the GSC theory with $a=0.9$.
}
\end{figure}

Finally, to understand the $N$-dependence of the coil--to--globule
transition in Fig.~\ref{fig:CAP} we present the plots of the
specific energy slope, $N^{-1}d{\cal E}^{(int)}/dU^{(0)}$, vs
$U^{(0)}$ for the flexible rings of different sizes.
These curves nearly coincide in the repulsive coil region,
starting to diverge from a value of $U^{(0)}\gtrsim 1.2$.
The region of the transition,
where the quantity $N^{-1}d{\cal E}^{(int)}/dU^{(0)}$ experiences
the most dramatic drop, becomes increasingly narrower with 
increasing polymer size $N$. Moreover, since the magnitude
of the overall change in the specific energy slope
also increases with $N$, the coil--to--globule transition
becomes `sharper' with $N$ in the GSC theory, consistent with
the MC simulation data and the transition being of second order
\cite{CloizeauxBook,Edwards}. 
Note also, that the theta--point,
which we may define e.g. as the point of the maximal
change in $N^{-1}d{\cal E}^{(int)}/dU^{(0)}$, shifts
towards lower values of $U^{(0)}$ with increasing $N$.

\section{Conclusion}\label{sec:concl}

In this paper we have developed a version of the Gaussian self--consistent 
(GSC) technique which does not rely on the virial--type expansion of the 
Hamiltonian in terms of powers of the density, 
$\int d {\bf r}\, \rho({\bf r})^L$. As a result, it is now possible to apply
the new method to practically any polymer model involving conventional 
molecular interactions.

Thus, we have been able to compare the mean spatial characteristics 
and the energy values between the results from the GSC theory and Monte Carlo
(MC) simulation based on precisely the same model, which includes 
the harmonic bonded
and the Lennard--Jones pair--wise interactions. This comparison has
been performed for three types of macromolecular architectures of
an isolated chain: a ring, an open polymer, and a star.
We have done this also across the range of the
coil--to--globule transition, as well as for different degrees of
polymerisation and degrees of flexibility.
Naturally, the GSC theory agrees with the second order
nature of the coil--to--globule transition for flexible homopolymers.

Importantly, the speed of numeric computation is much faster \cite{CPU}
in the GSC method than in the equivalent
MC simulation for obtaining the same data, 
particularly so for systems possessing an extra
kinematic symmetry, such as for rings or stars.

Overall, the agreement in the shapes of the curves and many of the
particular numerical values of observables between GSC and MC is better than 
one could have anticipated given the relative simplicity of the GSC
technique. Where any significant level of deviation does occur, it has been
identified as either related to the correlation hole effect
at small separations, or to the stretched Gaussian behaviour at large
separations, in the radial distribution function (RDF) \cite{CorFunc}.

In particular, for the repulsive coil, the energy and the mean--squared
distances between near monomers along the chain are quite accurate
in the GSC theory as compared to the MC data
(with a typical deviation of several percent), but the
distances between remote monomers, and hence the radius of gyration,
are overestimated by the GSC method (with a typical deviation of a few 
dozen percent).
This is a well known drawback of such a theory,
related to the overestimation of the Flory swelling exponent
for long chains, due to the fact that the RDF here does not have a 
stretched exponential behaviour at large separations.
On the contrary, for a rather stiff coil, the GSC theory
underestimates the radius of gyration. This may be due to 
shortcomings of the hard sphere Carnahan--Starling term.
For the collapsed globule, on the other hand, the distances
and the  radius of gyration are quite accurate in the GSC theory, 
although the mean energy
is less so because of the lack of a sharp liquid--like
peak in the RDF.

To make the agreement of GSC and MC better, one has to finally overcome
the most restrictive feature of the method --- the Gaussian shape
of RDF itself. One possible way of doing this is to take a linear 
superposition of the Gaussian trial functions, thereby permitting `stretching' 
of the Gaussoid. This should be sufficient for curing the problem with the
swelling exponent of the repulsive coil in the GSC theory. 
Work along these lines is currently in progress.
The main difficulties in doing this are in the considerable mathematical
complications when calculating the non--Gaussian conformational
entropy of the chain, as well as in the added numerical complexity, since
a radial mesh for RDF would have to be introduced.
Nevertheless, we intend to resolve these issues and hope to present a more
accurate, and at last a non--Gaussian self--consistent theory in the 
near future.

\acknowledgments

The authors are grateful for interesting discussions to
Professor F.~Ganazzoli, Professor H.~Orland, Dr G.~Raos, Dr T.~Garel,
and to R.~Connolly for his help. The support of the Enterprise Ireland
international collaboration grants IC/2001/074 and BC/2001/034, as well
as the IRCSET basic research grant SC/02/226 are also acknowledged.
 



\newpage
\section*{Figure Captions}

\newpage

\section*{Tables}

\begin{table}
\caption{\label{tab:1}
Comparison of the mean energy values ${\cal E}^{(int)}$
(in $k_B T$ units) for different homopolymers
from the MC simulation (second column) based on the data set
of Ref. \cite{CorFunc} and from the GSC theory with the parameter
$a=0.9$ (third column) and $a=1.0$ (fourth column).
The fifth and sixth columns contain the relative deviation
$\delta=({\cal E}^{(int)}(GSC) / {\cal E}(MC) - 1)\times 100 \%$
for these two cases given as a percentage. The model parameters,
which are suppressed within the tables, such as e.g. $\lambda$,
$U^{(0)}$, are equal to zero.
}
\vskip 5mm

\begin{tabular}{|l|c|cc|cc|}
System    &   MC  & GSC, $a = 0.9$ & GSC, $a = 1.0$ &
$\delta_{a=0.9},\ \%$ & $\delta_{a=1.0},\ \%$ \\
\hline
{\bf Ring, $N=100$} &     &          &         &           &           \\
$\kappa=0.01$  &  148.6   &  148.58  &  148.59 &    0      &    0      \\
$\kappa=0.1$   &  151.21  &  151.22  &  151.51 &    0      &   0.2     \\
$\kappa=0.2$   &  155.24  &  155.52  &  156.22 &   0.2     &   0.6     \\
$\kappa=0.5$   &  168.3   &  170.03  &  172.0  &   1.0     &   2.2     \\
$\kappa=2.0$   &  232.4   &  232.1   &  239.0  & $-0.1$    &   2.8     \\
\hline
{\bf Ring, $\kappa=1$} &   &    &         &           &           \\
$N= 50$   &   94.76  &  96.08   &  97.99  &   1.4     &   3.4     \\
$N=100$   &   190.4  &  193.1   &  196.9  &   1.4     &   3.4     \\
$N=200$   &   381.5  &  386.9   &  394.4  &   1.4     &   3.4     \\
$N=300$   &   572.6  &  580.6   &  591.9  &   1.4     &   3.4     \\
\hline
{\bf Open, $\kappa=1$}  &   &   &         &           &           \\
$N=150$   &   284.2  &  287.1   &  292.7  &   1.0     &   3.0     \\
$N=200$   &   378.7  &  383.9   &  391.3  &   1.4     &   3.3     \\
\hline
{\bf Semi--flexible Ring} &  &  &         &           &           \\
{\bf $\lambda=1,\,\kappa=1$} &  &  &      &           &           \\
$N= 50$   &   106.6  &  104.9   &  107.3  & $-1.6$    &   0.7     \\
$N=100$   &   213.3  &  210.2   &  215.1  & $-1.4$    &   0.8     \\
$N=200$   &   426.8  &  420.9   &  430.7  & $-1.4$    &   0.9     \\
$N=300$   &   640.2  &  631.6   &  646.2  & $-1.3$    &   0.9     \\
\hline
{\bf Semi--flexible Ring} &  &  &         &           &           \\
{\bf $\lambda=5,\,\kappa=1$} &  &   &     &           &           \\
$N= 50$   &   110.1  &  110.5   &  113.2  &   0.4     &   2.8     \\
$N=100$   &   218.6  &  220.5   &  225.7  &   0.9     &   3.2     \\
$N=200$   &   436.2  &  440.9   &  451.4  &   1.1     &   3.5     \\
$N=300$   &   653.9  &  661.4   &  677.0  &   1.1     &   3.5     \\
\hline
{\bf Star, $N/f=50,\,\kappa=1$} &  &  &   &           &           \\
$f= 3$    &   286.5  &  295.1   &  301.2  &   3.0     &   5.1     \\
$f= 6$    &   575.1  &  596.5   &  608.8  &   3.7     &   5.9     \\
$f= 9$    &   864.8  &  902.1   &  920.8  &   4.3     &   6.5     \\
$f=12$    &   1156.  &  1211.   &  1236.  &   4.7     &   6.9     \\
\hline
{\bf Globule of a Ring} &    &     &      &           &           \\
{\bf $U^{(0)} = 6,\,\kappa=1$} &  &   &   &           &           \\
$N=100$   & $-426.7$ & $-287.2$ & $-211.9$ & $-32.7$  &  $-50.3$  \\
$N=200$   & $-1000.$ & $-722.6$ & $-540.3$ & $-27.7$  &  $-46.0$  \\
\end{tabular}

\end{table}

\begin{table}
\caption{\label{tab:2}
Comparison of the mean--squared radius of gyration values 
$3{\cal R}_g^2$ (in $\ell^2$ units) for different homopolymers
from the MC simulation (second column) based on the data set
of Ref. \cite{CorFunc} and from the GSC theory with the parameter
$a=0.9$ (third column) and $a=1.0$ (fourth column).
The fifth and sixth columns contain the relative deviation
$\delta=({\cal R}_g^2(GSC) / {\cal R}_g^2(MC) - 1)\times 100 \%$
for these two cases given as a percentage.
}
\vskip 5mm

\begin{tabular}{|l|c|cc|cc|}
System    &   MC  & GSC, $a = 0.9$ & GSC, $a = 1.0$ &
$\delta_{a=0.9}, \ \%$ & $\delta_{a=1.0},\ \%$ \\
\hline
{\bf Ring, $N=100$} &   &       &         &           &           \\
$\kappa=0.01$ &    2497.  &   2503.6 & 2504.6  &   0.3     &   0.3     \\
$\kappa=0.1$  &    273.57 &   275.18 &  277.8  &   0.6     &   1.5     \\
$\kappa=0.2$  &    153.56 &   157.61 &  160.81 &   2.6     &   4.7     \\
$\kappa=0.5$  &     82.68 &    91.04 &   94.69 &  10.1     &  14.5     \\
$\kappa=2.0$  &     45.98 &    60.70 &   64.84 &  32.0     &  41.0     \\
\hline
{\bf Ring, $\kappa=1$}   &   &       &         &           &           \\
$N= 50$  &     26.66 &    28.90 &   30.32 &   8.4     &  13.7     \\
$N=100$  &     61.45 &    70.30 &   74.16 &  14.4     &  20.7     \\
$N=200$  &     141.4 &    172.3 &   182.7 &  21.8     &  29.2     \\
$N=300$  &     232.4 &    292.0 &   310.3 &  25.6     &  33.5     \\
\hline
{\bf Open, $\kappa=1$}  &   &   &         &           &           \\
$N=150$  &     185.7 &    221.4 &   234.3 &  19.2     &  26.2     \\
$N=200$  &     253.5 &    321.6 &   341.0 &  26.9     &  34.5     \\
\hline
{\bf Semi--flexible Ring} &  &  &         &           &           \\
{\bf $\lambda=1,\,\kappa=1$} &  &  &      &           &           \\
$N= 50$  &     37.91 &    38.18 &   40.71 &   0.7     &   7.4     \\
$N=100$  &     86.07 &    92.97 &   99.44 &   8.0     &  15.5     \\
$N=200$  &     193.6 &    222.9 &   239.0 &  15.1     &  23.5     \\
$N=300$  &     305.1 &    371.5 &   398.8 &  21.8     &  30.7     \\
\hline
{\bf Semi--flexible Ring} &  &  &         &           &           \\
{\bf $\lambda=5,\,\kappa=1$} &  &  &      &           &           \\
$N= 50$  &     78.05 &    60.60 &   65.58 & $-22.3$   & $-16.0$   \\
$N=100$  &     210.8 &    161.0 &   175.8 & $-23.6$   & $-16.6$   \\
$N=200$  &     499.8 &    382.5 &   419.0 & $-23.5$   & $-16.2$   \\
$N=300$  &     861.3 &    622.0 &   681.5 & $-27.8$   & $-20.8$   \\
\hline
{\bf Star, $N/f=50,\,\kappa=1$} &  &  &   &           &           \\
$f= 3$   &     146.2 &   176.4  &  186.6  &   20.6    &  27.6     \\
$f= 6$   &     185.2 &   239.0  &  253.1  &   29.0    &  36.7     \\
$f= 9$   &     208.7 &   278.0  &  294.5  &   33.2    &  41.1     \\
$f=12$   &     228.7 &   308.1  &  326.5  &   34.7    &  42.8     \\
\hline
{\bf Globule of a Ring} &    &     &      &           &           \\
{\bf $U^{(0)} = 6,\,\kappa=1$} &  &   &   &           &           \\
$N=100$  &   6.126   &   6.431  &  7.384  &    5.0    &  20.5     \\
$N=200$  &   9.368   &   9.872  &  11.31  &    5.4    &  20.7     \\
\end{tabular}

\end{table}


\begin{references}

\bibitem{ConfTra}
E.G. Timoshenko, Yu.A. Kuznetsov, K.A.~Dawson. 
{\it Phys. Rev.} {\bf E 57} 6801 (1998).

\bibitem{Ganazzoli-98}
F. Ganazzoli.
{\it J. Chem. Phys.} {\bf 108} 9924 (1998).

\bibitem{Ganazzoli-00}
F. Ganazzoli. 
{\it J. Chem. Phys.} {\bf 112} 1547 (2000).

\bibitem{Torus} 
Yu.A.~Kuznetsov, E.G.~Timoshenko.
{\it J. Chem. Phys.} {\bf 111} 3744 (1999).

\bibitem{Ganazzoli}
F.~Ganazzoli, R.~La~Ferla, G.~Allegra, 
{\it Macromol.} {\bf 28} 5285 (1995).

\bibitem{Netz}
R.R Netz, H. Orland.
{\it Eur. Phys. J.} {\bf 8} 81 (1999).

\bibitem{KineticLaws}
Yu.A. Kuznetsov, E.G. Timoshenko, K.A. Dawson.
{\it J. Chem. Phys.} {\bf 104} 3338 (1996).

\bibitem{GSC1st}
E.G. Timoshenko, Yu.A. Kuznetsov, K.A. Dawson.
{\it J. Chem. Phys.} {\bf 102} 1816 (1995).

\bibitem{CopStar}
F.~Ganazzoli, Yu.A.~Kuznetsov, E.G.~Timoshenko.
{\it Macromol. Theory Simul.} {\bf 10} 325 (2001).

\bibitem{Ganazzoli-dendr}
F.~Ganazzoli, R.~La~Ferla.
{\it J. Chem. Phys.} {\bf 113} 9288  (2000);
F.~Ganazzoli, R.~La~Ferla, G.~Raffaini.
{\it Macromol.} {\bf 34} 4222 (2001).

\bibitem{QzMess}
Yu.A. Kuznetsov, E.G. Timoshenko.
{\it Il Nuovo Cimento} {\bf 20D (12bis)} 2265 (1998).

\bibitem{Analysis}
E.G. Timoshenko, Yu.A. Kuznetsov.
{\it J. Chem. Phys.} {\bf 112} 8163 (2000).

\bibitem{Edwards}
M.~Doi, S.F.~Edwards. {\it The Theory of Polymer Dynamics}. Oxford Science.
New York (1989).

\bibitem{CloizeauxBook} J. des Cloizeaux, G.~Jannink.
{\it Polymers in Solution.} Oxford Science Publ (1990).

\bibitem{Edwards-var}
S.F. Edwards, P. Singh.
{\it J. Chem. Soc. Faraday Trans.} {\bf  II 75} 1001 (1979).

\bibitem{Cloizeaux-var}
J. des~Cloizeaux. 
{\it J. de Physique} {\bf 31} 715 (1970).

\bibitem{OrlandPitard}
E. Pitard, H. Orland.
{\it Europhys. Lett.} {\bf 41} 467 (1998).

\bibitem{Pitard}
E. Pitard,
{\it Eur. Phys. J.} {\bf  B 7} 665 (1999). 

\bibitem{Eu}
H.H.~Gan, B.C.~Eu.
{\it J. Chem. Phys.} {\bf 99} 4084 (1993); {\it ibid} 4103 (1993).

\bibitem{Taylor} 
M.P.~Taylor, J.E.G.~Lipson.
{\it J. Chem. Phys.} {\bf 104} 4835 (1996); {\bf 106} 5181 (1997).

\bibitem{Allegra}
 G.~Allegra, F.~Ganazzoli.
{\it J.~Chem.~Phys.} {\bf 83} 397 (1985).

\bibitem{Allegra-93}
G. Allegra, E. Colombo, F. Ganazzoli.
{\it Macromol.} {\bf 26} 330 (1993).

\bibitem{Raos-94}
G. Raos, G. Allegra, F. Ganazzoli.
{\it J. Chem. Phys.} {\bf 100} 7804 (1994).

\bibitem{CorFunc}
E.G. Timoshenko, Yu.A. Kuznetsov, R. Connolly. 
{\it J. Chem. Phys.} {\bf 116} 3905 (2002).

\bibitem{HansenMacDonald}
J.-P.~Hansen, I.R.~McDonald. {\it Theory of simple liquids}. 
Academic Press, London (1990).

\bibitem{CarnahanStarling}
N.F.~Carnahan, K.E.~Starling. 
{\it J. Chem. Phys.} {\bf 51} 635 (1969).

\bibitem{Talanquer}
V.~Talanquer, D.W.~Oxtoby. {\it Faraday Discuss.} {\bf 112} 91 (1999);
M.~Robles, M.L.~de~Haro.
{\it Phys. Chem. Chem. Phys.}
{\bf 3} 5528 (2001).

\bibitem{CombStar}
E.G. Timoshenko, Yu.A. Kuznetsov. 
{\it Colloids and Surfaces} {\bf A 190} 135 (2001).

\bibitem{CPU}
Benchmarks were performed on an AMD Athlon MP $1900+$ with the
following run time parameters. For the ring:
MC: $S=50,000$, $\Delta t=8\,N^2$, resulting in the
total run time $t_{total} \simeq 1000$ minutes; 
GSC: $n_{iterations} \simeq 5,000$, resulting in $t_{total} \lesssim 1$ minute. 
For the open chain:
MC: $S=50,000$, $\Delta t=40\,N^2$, resulting in 
$t_{total} \simeq 5,000$ minutes;
GSC: $n_{iterations}\simeq 18,000$, resulting in $t_{total} \simeq 25$ minutes.

\bibitem{Guida} R. Guida, J. Zinn--Justin.
{\it J. Phys.} {\bf A 31} 8103 (1998).

\end{references}
\end{document}